\documentclass[prl,twocolumn,showpacs,superscriptaddress,floatfix]{revtex4}
\usepackage{graphicx,color}

\newlength{\ldag}
\newcommand{\cra}{{a^\dagger}}
\newcommand{\ana}{{a^{\phantom\dagger}\hspace{-\ldag}}}
\newcommand{\crb}{{b^\dagger}}
\newcommand{\anb}{{b^{\phantom\dagger}\hspace{-\ldag}}}
\newcommand{\crc}{{c^\dagger}}
\newcommand{\anc}{{c^{\phantom\dagger}\hspace{-\ldag}}}
\newcommand{\crd}{{d^\dagger}}

\begin{document}

\title{Finite-Size Scaling Exponents in the Dicke Model}

\author{Julien Vidal}
\email{vidal@lptmc.jussieu.fr}
\affiliation{Laboratoire de Physique Th\'eorique de la Mati\`ere Condens\'ee, CNRS UMR 7600,
Universit\'e Pierre et Marie Curie, 4 Place Jussieu, 75252 Paris Cedex 05, France}

\author{S\'ebastien Dusuel}
\email{sdusuel@thp.uni-koeln.de}
\affiliation{Institut f\"ur Theoretische Physik, Universit\"at zu
K\"oln, Z\"ulpicher Str. 77, 50937 K\"oln, Germany}

\begin{abstract}
We consider the finite-size corrections in the Dicke model and determine the scaling exponents at the critical point for several quantities such as the ground state energy or the gap. Therefore, we use the Holstein-Primakoff representation of the angular momentum and introduce a nonlinear transformation to diagonalize the Hamiltonian in the normal phase. As already observed in several systems, these corrections turn out to be singular at  the transition point and thus lead to nontrivial exponents. We show that for the atomic observables, these exponents are the same as in the Lipkin-Meshkov-Glick model, in agreement with numerical results.  We also investigate the behavior of the order parameter related to the radiation mode and show that it is driven by the same scaling variable as the atomic one. 
\end{abstract}


\maketitle

%
%
%
%
Superradiance is the collective decay of an excited population of atoms {\it via} spontaneous emission of photons. This phenomenon first predicted by Dicke in 1954 \cite{Dicke54} has, since then,  been observed experimentally in several quantum optical as well as solid-state systems  (for a  review see Ref.~\cite{Brandes05_1}). The phase diagram of the Dicke model, which is the subject of the present study, has been established in the thermodynamical limit by Hepp and Lieb \cite{Hepp73} revealing the existence of a second-order quantum phase transition. 
This transition has been shown to be associated to a crossover between Poisson and Wigner-Dyson level statistics for a finite number of atoms $N$, thus raising the question of the finite-size corrections in this system \cite{Emary03_1,Emary03_2}. 
These corrections have also been shown to be crucial in the understanding of entanglement properties  \cite{Lambert04,Lambert05} which become trivial if one directly considers the thermodynamical limit \cite{Lambert04,Lambert05}. 
In these latter studies, nontrivial finite-size scaling exponents have been numerically found at the critical point and further been compared to those obtained in the Lipkin-Meshkov-Glick model \cite{Reslen05_1}. The aim of the present work is to determine these exponents.

To achieve this goal, we proceed in several steps. First, we use the Holstein-Primakoff boson representation \cite{Holstein40} for the atomic degrees of freedom which is well adapted for a  $1/N$ expansion of the Hamiltonian, $N$ being the number of atoms. Second, we exactly diagonalize the  expanded (quartic) Hamiltonian at order $1/N$. 
In a recent series of papers \cite{Dusuel04_3, Dusuel05_1, Dusuel05_3}, this diagonalization was performed using the Continuous Unitary Transformations (CUTs) methods \cite{Wegner01} but here, the problem is more complicated for several reasons: ({\it i}) it involves two different degrees of freedom; ({\it ii}) the parameter space is two-dimensional; and {\it (iii)} the total number of particle is not fixed. These complications render the analytical resolution of the flow equations coming from CUTs approach difficult \cite{Dusuel05_2}. 
We are thus led to use an alternative approach relying on a canonical transformation of the initial bosonic operators. This transformation provides both the eigenstates and the eigenspectrum of $H$, and thus allows one to compute any matrix element of any observable. Here, we focus on the quantities which have been numerically investigated and we show that their $1/N$ expansion is singular at the critical point. 
The analysis of these divergences directly provides the finite-size scaling exponents which are the same as in the Lipkin-Meshkov-Glick model, at least for the physical quantities involving atomic degrees of freedom. We also compute this exponent for the order parameter which is found to vanish as $N^{-2/3}$ at the transition point. 
Finally, we discuss numerical data which are in good agreement with our predictions.

%
%
%
%

Let us consider the single-mode Dicke Hamiltonian \cite{Dicke54} without the rotating wave approximation
%
%
\begin{equation}
   H=\omega_0 J_z+ \omega \cra \ana + {\lambda \over \sqrt{2j}} \left( \cra + \ana \right) \left(J_+ + J_-\right),
   \label{eq:hamiltonian}
\end{equation}
%
%
where $\cra$ and $\ana$ are bosonic creation and annihilation operators satifying $[\ana,\cra]=1$. The angular momentum operators are defined as  $J_{\alpha}=\sum_{i=1}^N \sigma_{\alpha}^{i}/2$ where the $\sigma_{\alpha}$'s are the Pauli matrices, and $J_\pm=J_x \pm i J_y$. 

This Hamiltonian, which describes the interaction of a photon field with $N$ two-level atoms (spins $1/2$),  conserves the magnitude $j$ of the pseudo-spin $\left(\left[H,{\bf J}^2 \right]=0\right)$. In the following, we focus on the sector $j=N/2$ to which the ground state belongs. 
 Further, one has $[H,\Pi]=0$ where 
%
%
\begin{equation}
   \Pi=e^{i \pi \left(\cra \ana +J_z+ j\right)},
\end{equation}
%
%
is the parity operator. An appropriate basis of the Hilbert space is thus provided by the states $|n\rangle \otimes |j,m\rangle$  where  $|n\rangle$ denotes an eigenstate of the photon density operator $\cra \ana$ with eigenvalue $n$, and $|j,m\rangle$ the eigenstate of ${\bf J}^2$ and $J_z$ associated to eigenvalues $j$ and $m$ respectively.
 
In the thermodynamical limit and at zero temperature, the system described by this Hamiltonian undergoes a second-order quantum phase transition at a critical couplings $\lambda_c=\sqrt{\omega \omega_0}/2$. 
As an order parameter of the transition, one can choose the expectation value of the photon number per atom in the ground state which satisfies:
%
%
\begin{equation}
\lim_{N \rightarrow \infty} \langle \cra \ana \rangle /N=\Bigg\{ 
\begin{array}{ccc}
0 & {\rm for}& \lambda < \lambda_c\\ 
{\lambda^2\over \omega^2}-{\omega_0^2 \over 16 \lambda^2} & {\rm for}&  \lambda \geq \lambda_c
\end{array}.
\end{equation}
%
%

As we shall see, nontrivial exponents are only found at the critical point that we will investigate from the normal (symmetric) phase, {\it i.e.}, for $\lambda < \lambda_c$. 
A convenient starting point to perform a $1/N$ expansion of the Hamiltonian is to use the Holstein-Primakoff boson representation of the angular momentum \cite{Holstein40} which reads:
%
%
\begin{eqnarray}
\label{eq:HP1}
J_+ &=&\crb \sqrt{N - \crb \anb}=\left(J_-\right)^\dagger,\\
\label{eq:HP2}
J_z &=& \crb \anb -{N \over 2}, 
\end{eqnarray}
%
%
with $[\anb,\crb]=1$, so that we now have to consider a two-boson problem. In the thermodynamical limit and for $\lambda < \lambda_c$, one has $ \langle \crb \anb \rangle /N \ll 1$ and we can expand the square root in (\ref{eq:HP1}) to obtain the following expanded form of the Hamiltonian:
%
%
\begin{eqnarray}
   H&=&-{N \over 2} \omega_0+\omega_0 \crb \anb+ \omega \cra \ana + \lambda \left( \cra+ \ana\right) \left( \crb+ \anb \right) \nonumber \\
&&-{\lambda \over 2 N} \left( \cra+ \ana \right) \left( \crb \anb^2 + \crb^2 \anb \right)
+O\left(1/N^2 \right). 
   \label{eq:hamiltonian_expanded}
\end{eqnarray}
%
%
Note that we restrict this expansion at the order $1/N$ which, as we will see thereafter, is sufficient for our purpose. At order $(1/N)^0$, the Hamiltonian is quadratic and can thus be diagonalized via a Bogoliubov transformation as already discussed in Ref.~\cite{Emary03_2}. The real problem arises at the order $1/N$ where one has to diagonalize a quartic form. 

%
%
%
%
As explained above, the CUTs formalism used in recent studies \cite{Dusuel04_3, Dusuel05_1, Dusuel05_3} for this step is difficult to implement in the Dicke model. Instead, we use here an approach that simply requires to solve a set of algebraic equations instead of differential equations. 
The main idea of this method is to perform the following canonical transformation  
%
%
\begin{eqnarray}
\cra&=&  \sum_{j= 0}^{p} {A^\dagger_j \over N^j},\\
\crb&=&   \sum_{j=0}^{p} {B^\dagger_j \over N^j},
\end{eqnarray}
%
%
where the $A^\dagger_p$ and $B^\dagger_p$ are polynomials functions of new bosonic operators $\crc, \anc,\crd,d$, such that $H$ expanded at order $1/N^p$ is a polynomial function in $n_c$ and $n_d$. 

At order zero, this transformation coincides with the Bogoliubov transformation and one has to determine 8 independent coefficients. Indeed, one has schematically: 
%
%
\begin{eqnarray}
A^\dagger_0&=&   \sum_{i,j,k,l} \alpha_{i,j,k,l}^{(0)}  \crc^i c^j \crd^k d^l,\\
B^\dagger_0&=&   \sum_{i,j,k,l} \beta_{i,j,k,l}^{(0)}  \crc^i c^j \crd^k d^l,
\end{eqnarray}
%
%
where $\alpha_{i,j,k,l}^{(q)}$ (resp. $\beta_{i,j,k,l}^{(q)}$) stands for the coefficient of $\crc^i c^j \crd^k d^l$ in the expansion of $A^\dagger_q$ (resp. $B^\dagger_q $). Since, at this order, the transformation is linear, the sum is constrained by $i+j+k+l=1$. The eight equations to be solved which are quadratic forms of the $\alpha's$ and $\beta$'s are, as usual, obtained by {\it (i)} requiring the cancellation of (nonconstant) terms which are not proportionnal to $n_c$ and $n_d$, and  {\it (ii)} imposing the following commutation rules,  
%
%
\begin{equation}
\left[\ana,\cra \right]=1,\left[\anb,\crb \right]=1, \left[\ana,\crb \right]=0, \left[\ana,\anb \right]=0.
\label{eq:comm}
\end{equation}
%
%
The full solution of these equations can be found in Ref.~\cite{Emary03_2}. 

Now, let us turn to the next order $p=1$ for which $H$ is quartic. At this order, the corresponding transformation reads
%
%
\begin{eqnarray}
A^\dagger_1&=&  \sum_{i,j,k,l} \alpha_{i,j,k,l}^{(1)}  \crc^i c^j \crd^k d^l, \label{eq:nonlinA} \\
B^\dagger_1&=&  \sum_{i,j,k,l}  \beta_{i,j,k,l}^{(1)}  \crc^i c^j \crd^k d^l, \label{eq:nonlinB}
 \end{eqnarray}
%
%
where the sum now contains two types of terms: linear ($i+j+k+l=1$) and cubic ($i+j+k+l=3$). There is thus $48$ independent parameters to be determined. At this order, these are the only terms that need to be present since the Hamiltonian (\ref{eq:hamiltonian_expanded}) only contains quadratic and quartic terms.
We also emphasize that once the $\alpha_{i,j,k,l}^{(0)}$'s and the $\beta_{i,j,k,l}^{(0)}$'s are known, the constraints to be satisfied are linear functions of the $\alpha_{i,j,k,l}^{(1)}$'s and  $\beta_{i,j,k,l}^{(1)}$'s. More generally, to determine the parameters  for $p\geq 1$, we must solve a set of linear equations involving only the $\alpha_{i,j,k,l}^{(q)}$'s with $q < p$. At order $p=1$, the equations to be solved are given by requiring the cancellation of (nonconstant) terms not proportionnal to $n_c$, $n_d$, $n_c^2$, $n_d^2$, and $n_c n_d$, but also by requiring the commutation rules (\ref{eq:comm}) to be satisfied.
Note that the spirit of this approach is the same as the one issued from the CUTs in which the running coupling, in the infinite time limit, identify with the $\alpha$ and $\beta$'s \cite{Dusuel05_2}.
%
%
%
%

The exact solutions of this set of equations are obviously too long to be given here, but let us sketch the main results that can be extracted from them. As already shown in several models \cite{Dusuel04_3, Dusuel05_1, Dusuel05_3}, the $1/N$ corrections to physical observables such as the gap or the order parameter display some singularities at the critical point. 
As detailed in \cite{Dusuel05_2}, the schematic form of an observable $\Phi$ in the vicinity of the critical point is:
%
\begin{equation}
  \Phi_N(\lambda)=
  \Phi_N^\mathrm{reg}(\lambda)+\Phi_N^\mathrm{sing}(\lambda),
\end{equation}
%
where the superscript reg and sing stands for regular and singular functions at $\lambda=\lambda_c$. By singular, we mean that the function and/or its derivatives with respect to $\lambda$ diverges at the critical point. Further, a close inspection of the $1/N$ expansion shows that near $\lambda_c$ one has:
%
\begin{equation}
  \Phi_N^\mathrm{sing}(\lambda)\simeq
  \frac{\Xi(\lambda)^{\xi_\Phi}}{N^{n_\Phi}}
  \mathcal{F}_\Phi\left[N\Xi(\lambda)^{3/2} \right],
  \label{eq:scaling}
\end{equation}
%
where $\Xi(\lambda)=\lambda_c-\lambda$ and $\mathcal{F}_\Phi$ is a function depending  only on the scaling variable $N\Xi(\lambda)^{3/2}$. The exponents $\xi_\Phi$ and $n_\Phi$ are characteristics of the observables $\Phi$. In the present study, we have only checked this scaling hypothesis at order $1/N$ but we strongly believe that, as in previous models we studied, one indeed has such a scaling variable. For instance, the ground state energy per atom near the critical point reads: 
%
\begin{equation}
 e_0 \simeq  c_0+ {1 \over N} \left[{c_1+ c_2 \Xi(\lambda)^{1/2}} \right]+{1 \over N^2} {c_3 \over  \Xi(\lambda)} +O\left(1/N^3 \right),
 \label{eq:gse}
\end{equation}
%
with 
%
%
\begin{eqnarray}
c_0&=& -\omega_0/2,\\
c_1&=&{1 \over 2} \left[-\omega-\omega_0 +(\omega^2+\omega_0^2)^{1/2} \right],\\
c_2&=&{(\omega \omega_0)^{3/4} \over (\omega^2+\omega_0^2)^{1/2}}, \\
c_3&=& {3 \omega^{5/2} \omega_0^{3/2} \over 64(\omega^2+\omega_0^2)}.
 \end{eqnarray}
%
%
Using the hypothesis (\ref{eq:scaling}), these expressions  allow us to identify  $\xi_{e_0}=1/2$ and $n_{e_0}=1$. Note that for the spectrum (only), one can also obtain these corrections by a standard first-order perturbation theory.
The most striking result is that the scaling variable $N\Xi(\lambda)^{3/2}$, which is the key ingredient for our study, does not depend on the observable. This remarkable fact already observed for single-boson model \cite{Dusuel04_3} is rather surprising here since one may have expected one different variable for each types of degrees of freedom. 
Furthermore, the Hamiltonian depends on two independent parameters but their value do not change the scaling variable. In particular, we find no difference between the resonant ($\omega=\omega_0$) and the off-resonant case.

To obtain the finite-size scaling exponent from the gene\-ral form (\ref{eq:scaling}) it is sufficient to underline that, at finite $N$, no divergence can occur in the behavior of the observables, even at the critical point. This straightforwardly implies that,  to cure the singularity coming from $\Xi(\lambda)^{\xi_\Phi}$, one must have $\mathcal{F}_\Phi (x) \sim x^{-2 \xi_\Phi /3}$. This behavior of $\mathcal{F}$ then leads to
$\Phi_N^\mathrm{sing}(x_\mathrm{c}) \sim N^{-(n_\Phi+2\xi_\Phi/3)}$. 
We have computed the finite-size scaling exponents for several observables which are summarized in Table \ref{tab:exponents}. For completeness, we also give the value of these quantities in the thermodynamical limit. 
%
%
\begin{table}[ht]
  \centering
  \begin{tabular}{|c|c|c|c|c|}
    \hline
    $\Phi$ & $\lim_{N \rightarrow \infty}$ &$\xi_\Phi$ & $n_\Phi$ & $-(n_\Phi+2\xi_\Phi/3)$\\
    \hline
    \hline
    $e_0$ & $-\omega_0/2$ & 1/2 & 1 & -4/3\\
    \hline
    $\Delta$ & 0 & 1/2 & 0 & -1/3\\
    \hline
    $\langle \cra \ana \rangle/N$ & 0 & -1/2 & 1 & -2/3\\
    \hline
    $2 \langle J_z \rangle/N$ & -1 & -1/2 & 1 & -2/3\\
    \hline
    $4 \langle J_z^2 \rangle/N^2$ & 1 & -1/2 & 1 & -2/3\\
    \hline
    $4 \langle J_y^2 \rangle/N^2$ & 0 & 1/2 & 1 & -4/3\\
    \hline
   $4 \langle J_x^2 \rangle/N^2$ & 0 & -1/2 & 1 & -2/3\\
    \hline
\end{tabular}
  \caption{Finite-size scaling exponents at the critical point for the ground state energy $e_0$, the gap  $\Delta$, the order parameter $\langle \cra a \rangle/N$. the magnetization per atom $\langle J_z \rangle/N$, and the two-point correlation function $\langle J_\alpha^2 \rangle/N^2$ for $\alpha=x,y,z$.}
  \label{tab:exponents}
\end{table}
%
%

 It is clear that the canonical transformations (\ref{eq:nonlinA}-\ref{eq:nonlinB}) we used to diagonalize the Hamiltonian at order $1/N$  allow us to compute any matrix element (not only diagonal) of any observable expressed in terms of the initial operators. Here, we only focused on ground state expectation values  (except for the gap) because these have already been numerically computed and can thus be directly checked.  
 
The finite-size scaling exponents at the critical point have been computed for three quantities \cite{Reslen05_1}: $\langle J_z \rangle/N$ $(-0.54 \pm 0.01)$, $\sqrt{\langle J_z^2} \rangle/N$ $(-0.35 \pm 0.01)$ and indirectly  $\langle J_y^2 \rangle/N$ $(-0.26 \pm 0.01)$.
These results are very close to our predictions which are $-2/3$, $-1/3$ and $-1/3$ respectively, as can be read in Table \ref{tab:exponents}. Nevertheless, it is true that  our results do not lie within the error bars proposed by Reslen {\it et al.}. The same discrepancy was already observed in the LMG model for which we have explicitely shown that it was due to the too small system sizes investigated \cite{Dusuel04_3,Dusuel05_2}. Here, we strongly believe that the asymptotic regime was also not reached but, unfortunately, it is difficult to consider significantly larger sizes as those studied in Ref.~\cite{Reslen05_1}. This clearly requires further numerical efforts \cite{Brandes05_2} which are beyond the scope of the present study.

Let us also mention that the concurrence $C$ studied in Ref. \cite{Lambert04} which measures the spin-spin entanglement \cite{Wootters98} reads 
%
\begin{equation}
 (N-1) C=1- 4 \langle J_y^2 \rangle/N.
 \end{equation}
%
We thus predict a finite-size scaling exponent for this (rescaled) concurrence which is 
$-1/3$. 

At first glance, these results are strikingly similar to those obtained in the LMG model \cite{Dusuel04_3,Dusuel05_2} and this calls for several comments. Indeed,  it is well-known that if one focuses on the atomic degrees of freedom, both systems are equivalent in the thermodynamical limit as shown with different methods \cite{Gibberd74,Brankov75,Reslen05_1,Liberti05_1}. However, the finite-size corrections fail to be captured through this mapping. For instance, in the Dicke model, one has $\lim_{N \rightarrow \infty} 4 \langle J_y^2 \rangle/N=\omega_0/(\omega^2+\omega_0^2)^{1/2}$ whereas it vanishes in the LMG model \cite{Dusuel04_3,Dusuel05_2}. 
Moreover, for the LMG model, these exponents were found to be related to the upper critical dimension and the mean-field critical exponents of the Ising model in a transverse magnetic field \cite{Botet83} which is the counterpart of the LMG model with short-range interactions. 
For the Dicke model, it is difficult to find such a mapping since one cannot simply consider it as a long-range interacting system which would admit a short-range equivalent. 
Consequently, the similarity between the exponents of these two models is a nontrivial result which shed light on a recent controverse on that subject \cite{Brankov05_1,Reslen05_2,Liberti05_2}. 

Unlike previous studies using CUTs, we have developed here an alternative simple perturbative approach relying on a canonical transformation which allows one to diagonalize the Hamiltonian at order $1/N$. 
This method can, in principle, be applied to many similar models involving more than one type of boson and requires to solve a set of linear equations. It is thus, {\it a priori} simpler than the CUTs technique even if the number of equations to be solved quickly grows with the order of the $1/N$ expansion. Whatever the approach chosen, the main result to keep in mind is that if one accepts the hypothesis of a unique scaling variable, it is sufficient to compute the first nontrivial correction of one observable (for example the ground state energy) to get all the exponents. Indeed, the determination of $\xi_\Phi$ and $n_\Phi$ for the other ones can already be infered from the quadratic approximation. 

Finally, let us quote a recent work \cite{Leyvraz05} where a semi-classical approach has been introduced to obtain the finite-size scaling exponent in the LMG model. It would be interesting to analyze the Dicke model within this framework to have a better understanding of the similarities between these two systems.

\acknowledgments

We wish to thank T. Brandes, B. Dou\c{c}ot, C. Emary, N. Lambert, J.-M. Maillard  and D. Mouhanna for fruitful discussions and valuable comments on the manuscript. 


\end{document}